\documentclass{article}

\usepackage{nicematrix}
\usepackage{arxiv}

\usepackage[utf8]{inputenc} 
\usepackage[T1]{fontenc}    
\usepackage{hyperref}       
\usepackage{url}            
\usepackage{booktabs}       
\usepackage{amsfonts}       
\usepackage{nicefrac}       
\usepackage{microtype}      
\usepackage{lipsum}
\usepackage{graphicx}
\graphicspath{ {./images/} }

\usepackage{mathtools}
\usepackage{multirow}
\usepackage{blindtext}
\usepackage{graphicx}
\usepackage{caption}
\usepackage{physics}
\usepackage{subcaption}
\usepackage{amssymb}
\usepackage{bbm}
\usepackage{amsmath}
\allowdisplaybreaks[1]
\usepackage{tikz}
\usetikzlibrary{shapes,positioning}
\usetikzlibrary{arrows.meta}
\usepackage{float}
\usepackage[utf8]{inputenc}
\usepackage{fancyhdr}
\usepackage{titlesec}
\usepackage{etoolbox}
\usepackage[titletoc]{appendix}
\usepackage{makeidx}
\usepackage{hyperref}
\usepackage{wrapfig}
\makeindex
\usepackage[nottoc]{tocbibind}

\newcommand\numberthis{\addtocounter{equation}{1}\tag{\theequation}}

\patchcmd{\chapter}{\thispagestyle{plain}}{\thispagestyle{fancy}}{}{}

\titleformat{\chapter}{\normalfont\huge\bf}{\thechapter.}{20pt}{\huge}

\newcommand{\beginsupplement}{%
        \setcounter{table}{0}
        \renewcommand{\thetable}{S\arabic{table}}%
        \setcounter{figure}{0}
        \renewcommand{\thefigure}{S\arabic{figure}}%
     }

\title{The time-dependent reproduction number for epidemics in heterogeneous populations}

\author{
 Ioana Bouros \\
  Department of Computer Science\\
  University of Oxford\\
  Oxford, UK \\
  \texttt{ioana.bouros@spc.ox.ac.uk} \\
   \And
 Robin N. Thompson \\
  Mathematical Institute\\
  University of Oxford\\
  Oxford, UK \\
  \texttt{robin.thompson@maths.ox.ac.uk } \\
  \And
 David Gavaghan \\
  Department of Computer Science\\
  University of Oxford\\
  Oxford, UK \\
  \texttt{david.gavaghan@dtc.ox.ac.uk } \\
  \And
  Ben Lambert \\
  Department of Statistics \& Pandemic Sciences Institute\\
  University of Oxford\\
  Oxford, UK \\
  \texttt{ben.lambert@stats.ox.ac.uk} \\
}

\begin{document}
\maketitle
\begin{abstract}
The time-dependent reproduction number $R_t$ can be used to track pathogen transmission and to assess the efficacy of interventions. This quantity can be estimated by fitting renewal equation models to time series of infectious disease case counts. These models almost invariably assume a homogeneous population. Individuals are assumed not to differ systematically in the rates at which they come into contact with others. It is also assumed that the typical time that elapses between one case and those it causes (known as the generation time distribution) does not differ across groups. But contact patterns are known to widely differ by age and according to other demographic groupings, and infection risk and transmission rates have been shown to vary across groups for a range of directly transmitted diseases. Here, we derive from first principles a renewal equation framework which accounts for these differences in transmission across groups. We use a generalisation of the classic McKendrick–von Foerster equation to handle populations structured into interacting groups. This system of partial differential equations allows us to derive a simple analytical expression for $R_t$ which involves only group-level contact patterns and infection risks. We show that the same expression emerges from both deterministic and stochastic discrete-time versions of the model and demonstrate via simulations that our $R_t$ expression governs the long-run fate of epidemics. Our renewal equation model provides a basis from which to account for more realistic, diverse populations in epidemiological models and opens the door to inferential approaches which use known group characteristics to estimate $R_t$.
\end{abstract}


\section{Introduction}
In the long run, an infectious disease epidemic grows if, on average, the number of new infections generated by each infectious individual exceeds one; contrastingly, it will subside if this value is below one. The (instantaneous) time-dependent reproduction number, $R_t$, is rigorously defined as the average number of secondary cases generated by an infected case at time $t$ assuming that transmission conditions remain the same in the future. $R_t$ is an emergent property of a pathogen spreading through a specific population, and it depends on the biology of the pathogen and the characteristics of the population, including any measures taken to limit its spread \cite{Flaxman(2020), Brauner(2021)}. Accordingly, determining $R_t$ is crucial for public health policymaking during epidemics.

Renewal equation models have, over the past two decades, become the predominant models used when $R_t$ is estimated from epidemiological time series, most commonly using time series of infectious disease case counts \cite{Fraser(2007), Cori(2013), Thompson(2019)}. Their success is owed to their relative simplicity: these models assume that new infections are caused by previous infections with time lags between parent and daughter infections; these time lags are assumed to be characterised by a \textit{generation time distribution} allowing for variation in this quantity across infector-infectee pairs \cite{Svensson(2007)}. In these models, the number of new infections caused by a typical new case at time $t$ is given by $R_t$. Usually, these models are fitted to infectious disease case counts (rather than infection counts, which are harder to observe) and, in so doing, $R_t$ is estimated. There is a large literature and range of software tools devoted to $R_t$ inference via renewal equations \cite{Fraser(2007), Cori(2013), Thompson(2019),Gostic(2020),Abbott(2020),Creswell(2023)}.

These inferential frameworks are built upon an implicit assumption of uniformity in the population, i.e.\ that disease transmission rates are the same across the population. This ignores what is widely known -- that disease transmission differs systematically across groups -- and explicitly modelled in other types of models of disease transmission dynamics. For example, compartmental models (e.g. \cite{Como}) are frequently structured by age, largely because, in many populations, the numbers and types of contact that individuals typically have with others depend strongly on their age. This is so widely recognised that there is a literature devoted to estimating so-called \textit{contact matrices} that capture this information (e.g. \cite{POLYMOD}). Infection risk can also differ according to other demographics \cite{Pijls(2021),Theodore(2023)}. The time period between one case and those it causes can also vary according to demographics: pathogens may undergo distinct dynamics within individuals from specific groups and be spread differently, and members of different groups may also adapt their behaviour to varying degrees if symptoms show. These differences can result in large variations between generation times across groups (e.g. for COVID-19, \cite{Kim(2022)}).

This shortcoming of the standard renewal framework has been recognised in previous work. It is possible to estimate a group-specific $R_t$ from infection time series by making assumptions about the rate at which pathogens spread within and between groups \cite{Glass(2021)}, although differences in generation time distributions according to group have not been considered; in such frameworks, an overall reproduction number can be determined through a weighted average of within- and between-group reproduction numbers. An alternative approach is to model the population using infection-age-structured partial differential equation (PDE) systems, with one PDE for each of the groups \cite{Green(2021)}. This approach follows the classic McKendrick–von Foerster equation framework which is central in the study of population dynamics in areas such as demography \cite{Mkendirk(1925),Foerster(1959),Murray(2002)}, where birth and death processes vary according to age, and we follow this approach here.

We suppose there are groups distinct in terms of their contact behaviour, infection risks and generation time distributions. Like \cite{Green(2021)}, we determine an analytical expression for $R_t$, but ours is simpler and does not involve the generation time distribution. Using a discrete-time version of the model (as most inferential routines for $R_t$ use discrete-time renewal models), we arrive at the same expression. We use simulation to demonstrate that our expression for $R_t$ behaves as expected: $R_t=1$ defines the boundary between long-run epidemic growth (if $R_t>1$) and long-run decline (if $R_t<1$). We show that this $R_t$ expression holds also for a stochastic version of our model, where, on average, the behaviour of these systems is delineated by the $R_t=1$ boundary. We also derive a relationship between $R_t$ and the calendar time growth rate of the epidemic, $r_t$, for a population organised into groups, and we show that $r_t$ is particularly sensitive to changes in the generation time of groups with the most contacts. Our framework naturally allows known characteristics of different groups to be incorporated into renewal equation models of epidemics and into associated estimates of epidemiological quantities.

\section{Methods} \label{Methods}
\subsection{Renewal equations used to infer $R_t$}
Renewal models are primarily used to estimate $R_t$, and we now describe the most basic, yet indicative, form of these models. This model is stochastic and discrete-time (typically with time-steps of one day) and assumes that the population is homogeneous and takes the form:
\begin{align}
\label{Uni-categorical Process}
    I(t) \sim \text{Poisson}(R_t\Lambda_t)\text{, where } \Lambda_t=\sum^{t-1}_{a=1}w_{a} I(t-a).
\end{align}
In this expression, $I(t)\geq 0$ represents counts of infections arising at time $t$; $\Lambda_t\geq 0$ is known as the \textit{transmission potential} at time $t$, which is a weighted sum of past infections where the weights are determined by a \textit{generation time distribution}. This distribution is a discrete probability distribution, $\{w_a\}_{a=1}^{\infty}$, such that $\sum^{\infty}_{a=1} w_{a}=1$, where $w_{a}\geq 0$ gives the probability that the time elapsing between a past infection and a daughter case is $a$ days.






\subsection{Derivation of the renewal equation for the McKendrick–von Foerster model}
\label{One group population}
Eq. \eqref{Uni-categorical Process} is a stochastic renewal equation that can be motivated by the classic McKendrick–von Foerster model \cite{Mkendirk(1925),Foerster(1959)},  which uses an age-structured partial differential equation to model population dynamics where there are birth and death processes that depend on age (e.g. \cite[chapter~1.7]{Murray(2002)}). This model is deterministic and, typically when used in the context of epidemics, models the numbers of those infected continuously and is continuous in time and age of infection. We walk through a derivation of the renewal equation for this classic model since our approach for modelling structured populations extends it. In \S\ref{sec:vonfoerster_homogeneous}, we make explicit our assumptions in using this framework to model epidemic dynamics for a homogeneous population.

Following closely the derivation and notation used in \cite[chapter~1.7]{{Murray(2002)}}, we denote the \textit{density} of infections still present at time $t$ which began $a$ days ago by $n(t,a)$; the total \textit{number} of infections is then given by $\int_0^{\infty} n(t,a)da$. Note that, in our version of this classic model, $a$ denotes the infection age and \textit{not} ages of infected individuals. In a small time increment $dt$, the conservation law for the population dictates that \cite[chapter~1.7]{Murray(2002)}:
\begin{equation}\label{eq:conservation_equation}
dn(t,a)=\frac{\partial n}{\partial t}dt + \frac{\partial n}{\partial a}da = \mu(a)dt,
\end{equation}
where $\mu(a)$ is the rate at which infections end. In our work, we do not keep track of the number of recovered or dead cases, essentially assuming that $\mu(a)=0$. Since infection age changes at the same rate as calendar time, $dt/da=1$, eq. \eqref{eq:conservation_equation} can then be simplified to the following partial differential equation (PDE):
\begin{align}
\label{part_diff_eq}
    \frac{\partial n}{\partial t} + \frac{\partial n}{\partial a} = 0.
\end{align}
We specify a boundary condition which dictates the rate at which new infections arise:
\begin{equation}\label{eq:renewal_abstract}
    n(t,0) = \int_0^\infty b(t,a) n(t,a) da,
\end{equation}
where $b(t,a)\geq 0,\forall t,a$ is the rate at which infections arising $a$ days ago generate new infections. In addition, we also assume that the birth rate
of new infections arising from very old infections is effectively zero, that is $b(t,a)=0,\forall t, \forall a \gg 0$. In what follows, we explicitly model only the population dynamics for $t\geq 0$, and, in order to close the system, we assume that at $t=0$, there is a density of infections given by
\begin{equation}\label{eq:zero_boundary}
    n(0,a) = f(a).
\end{equation}
We can consider two solution classes corresponding to distinct groups of infected individuals at time $t$, each of which has a different relationship between $a$ and $t$:
\begin{equation}\label{eq:characteristics}
a= \begin{cases}
t + a_0,\text{ for }a > t, \text{ i.e. for infections arising before $t=0$},\\
t-t_0,\text{ for }a \leq t, \text{ i.e. for infections arising from $t=0$ onwards},
\end{cases}
\end{equation}
where $a_0>0$ is the age of an infection at time $t=0$ for the first group, and $t_0$ is the time at which an infection arises for the second (where $0\leq t_0\leq t$). For each of the cases in eqs. \eqref{eq:characteristics}, we can write $n(t,a)=n(t,a(t)):=n(t)$, i.e. a function of $t$ only, which we can substitute into eq. \eqref{part_diff_eq} to yield an ordinary differential equation (ODE):
\begin{equation}\label{eq:characteristic_ode}
    \frac{dn}{dt} = 0,
\end{equation}
which is valid only along one of the so-called \textit{characteristic} lines defined in eqs. \eqref{eq:characteristics}. In this context, the characteristics correspond to infections arising at a particular point in time, $\tau$: if $a>t$, a unique value of $a_0$ gives $\tau<0$; if $a\leq t$, $\tau=t_0$.
For each infection onset time then eq. \ref{eq:characteristic_ode} has the solution:
\begin{equation}
    n(t,a(t)) = \text{const.}, 
\end{equation}
meaning that the number of individuals infected at a given time $t$ remains forever constant: this makes sense because, in our framework, we do not allow for cessation of an infection.

We now derive these constant population sizes for each of the cases in eq. \eqref{eq:characteristics} by considering these at conveniently chosen moments. When $a>t$, we can write $a=t+a_0$ meaning $n(t,t+a_0)=\text{const.}$, and when $t=0$, the population size is:
\begin{equation}\label{eq:a_above_t}
    n(0,a_0)=f(a_0)=f(a-t).
\end{equation}
When $a\leq t$, we can write $a=t-t_0$ and $n(t,t-t_0)=\text{const.}$, and, when $t=t_0$:
\begin{equation}\label{eq:t_above_a}
    n(t_0,0)=n(t-a,0).
\end{equation}
Collecting eqs. \eqref{eq:a_above_t} \& \eqref{eq:t_above_a}, we have the solutions for the population sizes for all infection onset times:
\begin{equation}\label{eq:system_solutions}
    n(t,a)= \begin{cases}
    f(a-t), a>t,\\
    n(t-a,0), a\leq t.
\end{cases}
\end{equation}
Substituting eqs. \eqref{eq:system_solutions} into eq. \eqref{eq:renewal_abstract} results in:
\begin{equation}\label{eq:nearly_renewal}
    n(t,0)=\underbrace{\int_0^{t} b(t,a) n(t-a,0) da}_{a \leq t} + \underbrace{\int_t^{\infty} b(t,a) f(a-t) da}_{a > t}.
\end{equation}
To avoid having to specify the initial conditions at $t=0$, we generally assume that $t$ is large such that $b(t,a')\approx 0, \forall a'>t$. With this eq. \eqref{eq:nearly_renewal} reduces to:
\begin{equation}\label{eq:nearly_renewal1}
    n(t,0)=\int_0^{t} b(t,a) n(t-a,0) da.
\end{equation}
Eq. \eqref{eq:nearly_renewal1} gives a recurrence relation for our system solution, which is an implicit equation for $n(t,0)$. In \S\ref{sec:vonfoerster_homogeneous}, we show how this can be explicitly solved.

We now define $I(t):=n(t,0)$ to represent a measure of new infections arising at time $t$, and eq. \eqref{eq:nearly_renewal1} then becomes:
\begin{equation}\label{eq:renewal_continuous}
    I(t) = \int_0^{t} b(t,a) I(t-a) da.
\end{equation}
Eq. \eqref{eq:renewal_continuous} is a continuous and deterministic renewal equation.

\subsection{The McKendrick–von Foerster model of an epidemic in a homogeneous population}\label{sec:vonfoerster_homogeneous}
We now make explicit our assumptions around how infections arise in a homogeneous population and adapt eq. \eqref{eq:renewal_continuous} accordingly. We then derive long-term solutions for this model.
We suppose that there are two distinct time-dependent processes that modulate the birth rate of new infections: (i) the total number of individuals an infected person comes into contact with over the course of their infectious period if their infection debuted at time $t$, $C(t)\geq 0$; and, (ii), the probability, $0\leq \gamma(t)\leq 1$, that each contact results in an infection; this implicitly assumes that individuals do not change their behaviour as their infection progresses; for example, it assumes that when symptoms appear infected individuals do not change how they socialise. We assume that there is also an infection-age-dependent process which governs how much time typically elapses between one infection starting and those it causes: we use a continuous probability distribution for this, with probability density function, $w(a)$. The birth rate of new infections is then given by: $b(t,a)=C(t)\gamma(t) w(a)$, and we can substitute this into eq. \eqref{eq:renewal_abstract} to produce a renewal equation for new infections:
\begin{equation}\label{eq:renewal_epidemic}
I(t) = \int_0^t C(t) \gamma(t)  w(a) I(t-a)da.
\end{equation}
The rate at which new infections are generated from past infections is, in this model, not explicitly contingent on the numbers of those infected in the past. This differs from, say, an SIR compartmental model, where the rate at which new infections are generated from past infections depends on the availability of susceptible individuals which diminishes as the epidemic spreads through a population. In addition, we do not consider the termination of an infection; that is the death rate of new infections is assumed $\mu(a)=0$.

Because eq. \eqref{eq:renewal_epidemic} lacks self-regulation, the numbers of new infections arising can only either grow or decline exponentially over time; there is no non-zero infection equilibrium. We then look for a similarity solution \cite[chapter~1.7]{Murray(2002)} of our system (for large $t$) of the form:
\begin{equation}\label{eq:similarity_soln}
    n(t,a)= \exp(rt) p(a),
\end{equation}
where $r$ is the growth rate of the epidemic (which can be negative if there is epidemic decline), and $p(a)$ is an age-dependent distribution. Substituting eq. \eqref{eq:similarity_soln} into the $a\leq t$ solution in eq. \eqref{eq:system_solutions} which has $n(t,a)=n(t-a,0)$ gives:
\begin{equation}
    p(a) = p(0) \exp(-ra),
\end{equation}
which we substitute back into eq. \eqref{eq:similarity_soln} to give us the solution:
\begin{equation}
    n(t,a) = p(0) \exp(r(t-a)).
\end{equation}
Consequently, the number of new infections grows (or declines) exponentially over time:
\begin{equation}\label{eq:incidence_exponential}
    I(t)=p(0)\exp(rt),
\end{equation}
as desired for our intended solution. Substituting this into eq. \eqref{eq:renewal_epidemic} and dividing both sides by common terms gives:
\begin{equation}\label{eq:euler_lotka}
    1 = \int_0^t C(t)\gamma(t) w(a) e^{- r a} da,
\end{equation}
which is known as the Euler-Lotka equation and was first derived by Lotka at the start of the 20th century; this extended a more specific result obtained by Euler in the 1700s \cite{Bacaer(2007)}.

Whether the infected population grows of declines exponentially and the rate at which it does so is determined by the value of $r$ which solves eq. \eqref{eq:euler_lotka}. There are no simple closed-form solutions to eq. \eqref{eq:euler_lotka}; instead, we show how the sign of $r$ is determined by particular terms in this equation.

\begin{figure}[H]
\centering
\includegraphics[width=0.9\linewidth]{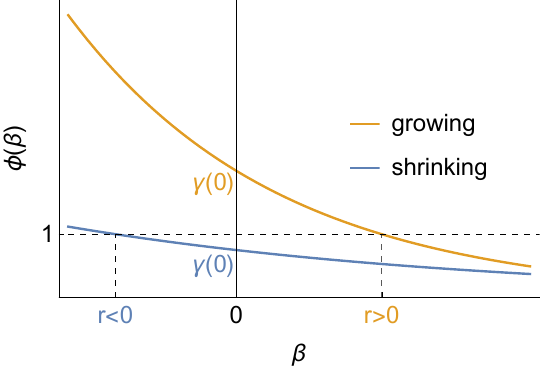}
\caption{\textbf{How the growth rate of an epidemic is determined by $\phi(0)$.} Here, $\phi(\beta) := \int_0^t C(t)\gamma(t) w(a) e^{- \beta a} da$. This figure is a reproduction of a figure from \cite[chapter~1.7]{Murray(2002)}.\label{fig:murray}}
\end{figure}

We define the function $\phi(\beta) := \int_0^t C(t)\gamma(t) w(a) e^{- \beta a} da$; we note that, using this notation, $r:=\{\beta, \text{ s.t. } \phi(\beta)=1\}$. Crucially, this function is monotonically decreasing in $\beta$. This means that if $\phi(0)>1$, then $r > 0$; if $\phi(0)=1$, $r=0$; and if $\phi(0)<1$, $r < 0$ (see Figure \ref{fig:murray}).

The sign of $r$ is so determined by the value of $\phi(0)$:
\begin{align}
    \phi(0) &= \int_0^t C(t)\gamma(t) w(a) da,\\
    &= C(t)\gamma(t) \int_0^t w(a) da,\\
    &\approx C(t)\gamma(t), \text{ for large $t$}.
\end{align}
Specifically, the population grows if $C(t)\gamma(t)>1$, is maintained if $C(t)\gamma(t)=1$ and shrinks if $C(t)\gamma(t)<1$. This behaviour is exactly the type of thresholding behaviour that is expected from a reproduction number, so we define:
\begin{equation}
    R(t) := C(t)\gamma(t).
\end{equation}
This time-varying reproduction number acts intuitively: $C(t)\gamma(t)$ is a measure of infections generated over the course of an infection (if transmission remains the same). If it exceeds 1, then each infection more than replaces itself; and analogously for the other two cases.

Formally, the reproduction number is defined as the largest (and here only) eigenvalue of the \textit{next-generation operator} \cite{Diekmann(2000)}; this is here an integral operator, $\mathcal{K}$:
\begin{equation}\label{eq:integral_operator}
    \mathcal{K}(g(a)):=\int_0^t C(t)\gamma(t) w(a) g(a) da.
\end{equation}
If $g(a)=1$, eq. \eqref{eq:integral_operator} yields the expected number of infections generated by an infection, i.e.
\begin{equation}
    \mathcal{K}(1)=C(t)\gamma(t),
\end{equation}
meaning $R(t)=C(t)\gamma(t)$ as required. In practice, $R(t)$ is written as $R_t$.

\subsection{Reproducing our simulation results}
Throughout this manuscript, we use simulations of unfolding outbreaks to support our mathematical derivations. To allow others to rerun these simulations, we make our Python code available through a public GitHub repository \cite{branchpro}.

\section{Results}
\subsection{A renewal equation for new infections in a structured McKendrick–von Foerster model}
\label{Multiple group population}
We now extend the model presented in \S\ref{Methods} and assume that the population is structured into interacting groups. In this section, we derive the renewal equation for new infections in this model.

We assume that the groups within the population differ in two ways that affect disease transmission:

\paragraph{Differences in contact patterns.} We take inspiration from compartmental ODE models where contact patterns within and between groups are incorporated through contact matrices (e.g. \cite{POLYMOD}).

\paragraph{Differences in within-host pathogen dynamics and behaviour during infection.} Within-person pathogen dynamics and transmission patterns can differ across individuals. Individuals can also differ in terms of their behaviour throughout infection: when symptoms show, more risk-averse individuals may choose to avoid socialising to prevent infecting others. Each of these characteristics could differ systematically across, for example, demographic groups, leading to differences in the generation time across these groups.

We define $n^i(t, a)$ as the density of current infections in population group $i\in\{1,2,...,N\}$ present at time $t$ that arose $a$ time periods ago. We can stack these densities into a vector containing this information for all groups:
\begin{align*}
    \underline{n}(t, a)= \begin{pmatrix}
n^1(t, a) \\
\vdots \\
n^N(t, a)
\end{pmatrix}.
\end{align*}
As for the homogeneous population model, applying population conservation yields the following system of PDEs:
\begin{align}
\label{multi_part_diff_eq}
    \frac{\partial \underline{n}}{\partial t} + \frac{\partial \underline{n}}{\partial a} = 0,
\end{align}
where $\partial \underline{n} /\partial x:=[\partial n^1 /\partial x,\partial n^2 /\partial x,...,\partial n^N /\partial x]'$ is the vector-partial derivative. 
We denote the birth rate of infections in group $i$ which were caused by infections in group $j$ by $b^{j \rightarrow i}(t,a)$. This means that the total density of new infections in group $i$ at time $t$ is given by those caused by infections across all groups:
\begin{align}
\label{Multiple Group Continous 1 group}
    n^i(t, 0) = \sum_{j=1}^N \int_0^\infty b^{j \rightarrow i}(a) n^j(t, a) da.
\end{align}
As for the homogeneous model (\S \ref{sec:vonfoerster_homogeneous}), we assume that the birth rate of new infections is the product of three terms:
\begin{equation}\label{eq:explicit_heterogeneous_birth}
    b^{j \rightarrow i}(t,a) = \gamma_t w^j(a) C^{(ji)}_t,
\end{equation}
where $C^{(ji)}_t$ now represents an element from a contact matrix and indicates the total number of contacts in group $i$ made by an individual in group $j$ over the course of their infection; $w^j(a)$ indicates a probability density function representing the generation time of group $j$. We make the additional assumption that the
probability that each contact results in an infection $\gamma_t$ is the same over all combinations of population groups. An alternative way to view this assumption is when setting $\gamma_t=1$ meaning $C^{(ji)}_t$ represents the total number of infections caused by group $i$ in those of category $j$ throughout their infection -- this implicitly assumes that the contact matrix accounts for variations in infectiousness and susceptibility across infector/infectee groups.

We can then use eqs. \eqref{Multiple Group Continous 1 group} \& \eqref{eq:explicit_heterogeneous_birth} to write an expression giving the rate of new infection generation across all groups:
\begin{align*}
    \underline{n}(t, 0) &= \begin{pmatrix}
\sum_{j=1}^N \int_0^\infty \gamma_t   w^j(a) C^{(1j)}_t n^j(t, a) da \\
\vdots \\
\sum_{j=1}^N \int_0^\infty \gamma_t  w^j(a) C^{(Nj)}_t n^j(t, a) da \\
\end{pmatrix} \\
&= \int_0^\infty \gamma_t  \begin{pmatrix}
\sum_{j=1}^N C^{(1j)}_t  w^j(a) n^j(t, a) \\
\vdots \\
\sum_{j=1}^N C^{(Nj)}_t  w^j(a) n^j(t, a) \\
\end{pmatrix} da\\
&= \int_0^\infty \gamma_t C_t \begin{pmatrix}
 w^1(a) n^1(t, a) \\
\vdots \\
 w^N(a) n^N(t, a) \\
\end{pmatrix} da\\
&= \int_0^\infty \gamma_t  C_t  \underbrace{\begin{bmatrix}
    w^1(a) & & \\
    & \ddots & \\
    & & w^N(a)
  \end{bmatrix}}_{W(a)} \begin{pmatrix}
 n^1(t, a) \\
\vdots \\
 n^N(t, a) \\
\end{pmatrix} da\\
&= \int_0^\infty \gamma_t  C_t W(a) \begin{pmatrix}
 n^1(t, a) \\
\vdots \\
 n^N(t, a) \\
\end{pmatrix} da.
\numberthis \label{N_dim_eq}
\end{align*}
As $\int_0^\infty w^i(a)da=1, \forall i$, the matrix $W(a)$ satisfies the following property:
\begin{equation}
\int_0^\infty W(a)da = \mathbb{I}_N.
\end{equation}
In order to close the system, we assume that at $t=0$, there is a density of infections given by
\begin{equation}\label{eq:zero_boundary_vector}
    \underline{n}(0,a) = \underline{f}(a).
\end{equation}
We use the same approach as for the homogeneous population model to arrive at the analogous form of eq. \eqref{eq:nearly_renewal} for the structured population model:
\begin{equation}\label{eq:nearly_renewal_vectorised}
    \underline{n}(t,0) = \int_0^\infty \gamma_t  C_t W(a) \underline{n}(t,a) da = \int_0^t \gamma_t  C_t W(a) \underline{n}(t-a,0) da + \int_t^\infty \gamma_t  C_t W(a) \underline{f}(t-a) da.
\end{equation}
As before, we assume that $t$ is large such that infections with age exceeding this time contribute negligibly to onward transmission, and eq. \eqref{eq:nearly_renewal_vectorised} then reduces to:
\begin{equation}\label{eq:nearly_renewal_vectorised1}
    \underline{n}(t,0) = \int_0^t \gamma_t  C_t W(a) \underline{n}(t-a,0) da,
\end{equation}
which can be written as a renewal equation for new infections:
\begin{equation}\label{eq:renewal_vectorised}
    \underline{I}(t) = \int_0^t \gamma_t  C_t W(a) \underline{I}(t-a) da.
\end{equation}

\subsection{The time-varying reproduction number for structured populations}\label{sec:time_varying_structured_Rt}
The multiple group renewal equation, eq. \eqref{eq:renewal_vectorised}, appears of the same form as the equivalent expression for a single group, eq. \eqref{eq:renewal_continuous}. It is natural then to suppose that the long-term solution for the multiple group model should be of analogous form:
\begin{equation}\label{eq:multiple_group_solution}
\underline{I}(t) = e^{r t} \underline{\Phi},
\end{equation}
where $\underline{\Phi}$ is a constant non-negative vector.

Counterintuitively, eq. \eqref{eq:multiple_group_solution} is true only under certain conditions.

To give a counterexample, suppose that the total contact matrix, $C_t$, is diagonal, meaning that the groups do not interact. In this case, it is not generally true that the growth rate of new infections should be identical across groups, as eq. \eqref{eq:multiple_group_solution} supposes. For example, if groups have differing generation times and they do not interact, the infections will generally grow at different calendar rates in each group.

In \S\ref{sec:thieme_renewal}, we outline the set of conditions for eq. \eqref{eq:multiple_group_solution} to hold. We then prove that these conditions hold in general for realistic population structures where groups interact; the only exception are populations in which not all population groups mix (as seen in eq. \eqref{non-mixing Ct}) and diseases where the generation-time interval $w$ has only one non-zero entry (that is subsequent infections occur exactly $j$ days after the parent infection for $w(j) \neq 0$).

We demonstrate that eq. \eqref{eq:multiple_group_solution} holds using simulations of a discrete deterministic renewal model; we set parameters of the model such that there is epidemic growth. In these simulations, we assume that the population is structured into three age groups: 0-20 year-olds, 20-65 year-olds and those aged $65+$. We use a contact matrix derived from \cite{POLYMOD} as proxy for the total contact matrix for this population:
\begin{equation}\label{eq:contact_matrix}
    C_t = \begin{pmatrix}
6.33 & 1.61 & 0.1\\
4.65 & 7.41 & 0.51\\
0.49 & 1.17 & 1.13
\end{pmatrix},
\end{equation}

where the rows correspond to infectors aged 0-20 (first row), 20-65 (second row) and $65+$ (third row); the columns correspond to infectees in groups with the corresponding ordering as for the rows. The values in the matrix represent the average daily contacts an individual in the row age group has with those in the column age group. Those aged 0-20 have the vast majority of their contacts with members of the same group; those aged 20-65 have high contact rates with the two youngest age groups; those aged 65+ have comparably low contact rates across all age groups.

\begin{figure}[H]
\centering
\includegraphics[width=0.9\linewidth]{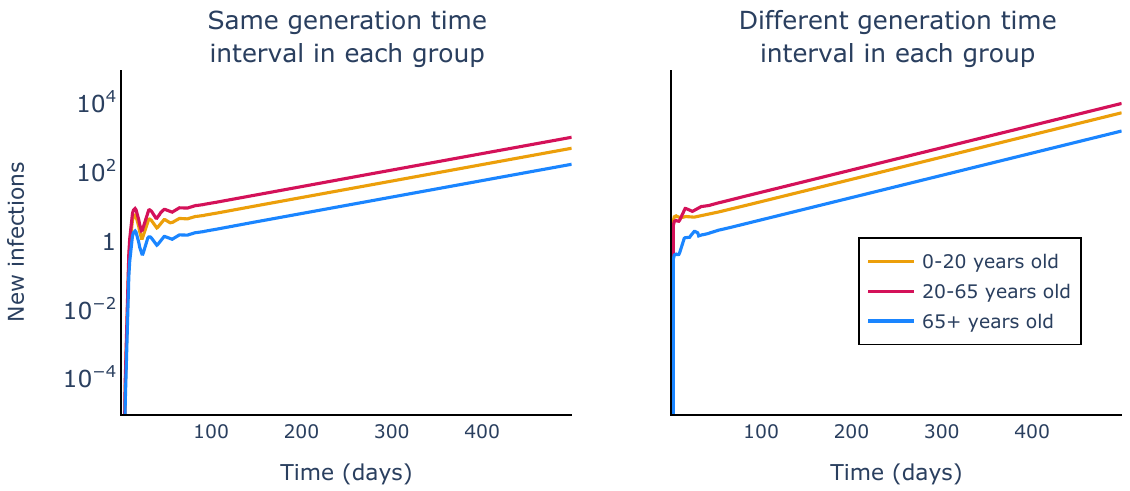}
\caption{\textbf{Epidemic growth occurs at the same rate across interacting groups.} For both panels, we use the same reproduction number $R_t=1.2$ and simulate using a discrete deterministic renewal equation. In the left-hand panel, we assume the generation time distributions differ across groups and are given by scenario 1 in Table \ref{tab:age_group_generation_times}; in the right-hand panel, the generation time distributions are given by scenario 2 of the same table. The contact matrix assumed is given in eq. \eqref{eq:contact_matrix}.
\label{Proof of concept}}
\end{figure}

In Figure \ref{Proof of concept} we show the results of these simulations; the right-hand panel shows the evolution of the outbreaks in each group when the groups all have the same generation time distribution; the left-hand panel shows the equivalent when the generation time distributions differ across groups. In both cases, the rate of outbreak growth is the same across the three populations, as is predicted by eq. \eqref{eq:multiple_group_solution}.

\begin{table}[]
    \centering
\renewcommand{\arraystretch}{1.3}
\begin{tabular}{|>{\centering\arraybackslash}m{2cm}|>{\centering\arraybackslash}m{2cm}|>{\centering\arraybackslash}m{2cm}|>{\centering\arraybackslash}m{2cm}|>{\centering\arraybackslash}m{2cm}|>{\centering\arraybackslash}m{2cm}|>{\centering\arraybackslash}m{2cm}|}
\hline
         & \multicolumn{3}{c|}{\textbf{Mean generation time}} &   \multicolumn{3}{c|}{\textbf{Standard deviation generation time}} \\
\hline
\textbf{Scenario} & 0-20 & 20-65 & 65+ & 0-20 & 20-65 & 65+ \\
\hline
1 & 5.6 & 15.3 & 25 & 3.3 & 3.3 & 3.3 \\
2 & 15.3 & 15.3 & 15.3 & 3.3 & 3.3 & 3.3 \\
3 & 25 & 15.3 & 5.6 & 3.3 & 3.3 & 3.3 \\
4 & 5 & 5.3 & 5 & 3.3 & 3.3 & 3.3 \\
5 & 5.1 & 5.1 & 5.1 & 3.3 & 3.3 & 3.3 \\
6 & 150.3 & 7 & 25 & 9.3 & 5 & 10 \\
7 & 30 & 25.3 & 15 & 3.3 & 3.3 & 3.3 \\
\hline
\end{tabular}
\caption{\textbf{Generation time distributions across age groups used in our simulation scenarios.}}\label{tab:age_group_generation_times}
\end{table}

Because eq. \eqref{eq:multiple_group_solution} holds for realistic populations, we assume a solution of this form and substitute it into eq. \eqref{eq:renewal_vectorised}. Simplifying both sides, the resulting equation reads as
\begin{align}
 \underline{\Phi} = \Big(\int_0^\infty \gamma_t  C_t W(a) e^{-r a} da\Big) \underline{\Phi} := \overline{K}(r) \underline{\Phi},
 \label{Complex Criterion}
\end{align}
\noindent where $\overline{K}(\beta) := \Big(\int_0^\infty \gamma_t  C_t W(a) e^{-\beta a} da\Big)$ is the Laplace integral transform of the positive matrix $\gamma_t  C_t W(a)$. Eq. \eqref{Complex Criterion} implies that the vector $\underline{\Phi}$ is an eigenvector of the positive matrix $\int_0^\infty \gamma_t  C_t W(a) e^{-r a} da$, with associated eigenvalue $1$.

We now use a linear algebra result called the \textit{Perron-Frobenius} theorem \cite{Bacaer(2023)} to convert eq. \eqref{Complex Criterion} into a form to which we can apply the same logic as for the single population model in \S\ref{sec:vonfoerster_homogeneous}. This theorem states that for a positive matrix, in our case $\overline{K}(\beta)$, the following three results hold:

\begin{enumerate}
    \item defining the eigenvalues of $\overline{K}(\beta)$ as $\nu_1,\nu_2,...,\nu_N$, the spectral radius $\rho(\overline{K}(\beta)):=\max(\{|\nu_1|,|\nu_2|,...,|\nu_N|\})$, is an eigenvalue of $\overline{K}(\beta)$;
    \item $\rho(\overline{K}(\beta)) \in \mathbb{R}$ and $\rho(\overline{K}(\beta)) > 0$;
    \item the eigenvector $\underline{v}$ of $\overline{K}(\beta)$ associated to the spectral radius $\rho(\overline{K}(\beta)$ is a positive vector and the one and only non-negative eigenvector of $\overline{K}(\beta)$ (all other non-negative eigenvectors are multiples of $\underline{v}$).
\end{enumerate}

All values of the vector $\underline{I}(t)$ are non-negative meaning that $\underline{\Phi}$ must also be a non-negative vector. Because $\underline{\Phi}$ is an eigenvector of $\overline{K}(r)$ corresponding to an eigenvalue of $1$. Then condition $3$ in the Perron-Frobenius theorem indicates that the spectral radius of $\overline{K}(r)$ is equal to $1$, i.e.
\begin{align}
    1=\rho(\overline{K}(r)).
    \label{Phi multiple group}
\end{align}
Eq. \eqref{Phi multiple group} is of a similar form to eq. \eqref{eq:euler_lotka} for the one group model. Like the right-hand side of eq. \eqref{eq:euler_lotka}, the function $\rho(\overline{K}(r))$, the spectral radius of the Laplace integral transform of a positive matrix, is also monotonically decreasing in $r$ \cite{Thieme(1984)} -- this means that we can apply the same reasoning as for the single-group model.

Since $\overline{K}(0)$ is the next generation operator \cite{Diekmann(2000)} yielding the expected numbers of newly infected in each of the groups assuming a single infected individual in each group, we define the overall reproduction number as:
\begin{align*}
R_t &= \rho(\overline{K}(0))\\
&= \rho \Big(\int_0^\infty \gamma_t  C_t W(a) e^{0 a} da\Big)\\
&= \rho(\gamma_t  C_t \int_0^\infty  W(a) da)\\
&= \rho(\gamma_t  C_t)\\
&= \gamma_t  \rho(C_t).
\end{align*}
Since the spectral radius of the Laplace integral transform $\overline{K}(r)$ of a positive matrix is monotonically decreasing \cite{Thieme(1984)}, we find that:
\begin{align}
\text{Epidemic growth} \iff r>0 \iff \rho(\overline{K}(0)) > \rho(\overline{K}(r)) \iff R_t = \gamma_t  \rho(C_t) > 1,
\label{Stability Criterion}
\end{align}
and similarly so with $R_t<1$ indicating long-term outbreak declines. We then define the time-varying reproduction number for a structured population, $R_t$, by
\begin{equation}\label{eq:rt_threshold}
    R_t = \gamma_t  \rho(C_t).
\end{equation}
As discussed in \S\ref{Methods}, renewal equations when used in practice are usually discrete; in \S\ref{sec:discrete_renewal_theory}, we show that the same expression for $R_t$ arises from the analogous discrete renewal model for a structured population. We use this discrete renewal model in the majority of our simulations, which is given by:
\begin{align}\label{eq:discrete_renewal_multigroup}
     I_t^{(j)} &= \frac{R_t}{\rho(C_t)} \sum^{N}_{i=1}C_t^{(ji)}\sum^{t-1}_{a=1}w_a^{(i)} I_{t-a}^{(i)},
\end{align}
where $C_t^{(ji)}$ represents the $i$-$j$th element of a matrix of total contacts made over the course of an infection started at time $t$; $\rho(C_t)$ is the largest eigenvalue of this contact matrix; $\{w_a^{(i)}\}$ represent elements of a discrete generation time interval distribution for group $i$.

\subsection{The long-run epidemic fate does not depend on the generation time interval}
Eq. \eqref{eq:rt_threshold} indicates that the reproduction number does not depend on the generation times of individual groups, which is a generalisation of results from previous work \cite{Green(2021)}. This makes intuitive sense because, in the long-run, an epidemic grows only if each infected case on average causes more than one subsequent infection -- it does not depend on \textit{when} those infections occur, which is what the generation time encodes.
\begin{figure}[ht]
\centering
\includegraphics[width=1\linewidth]{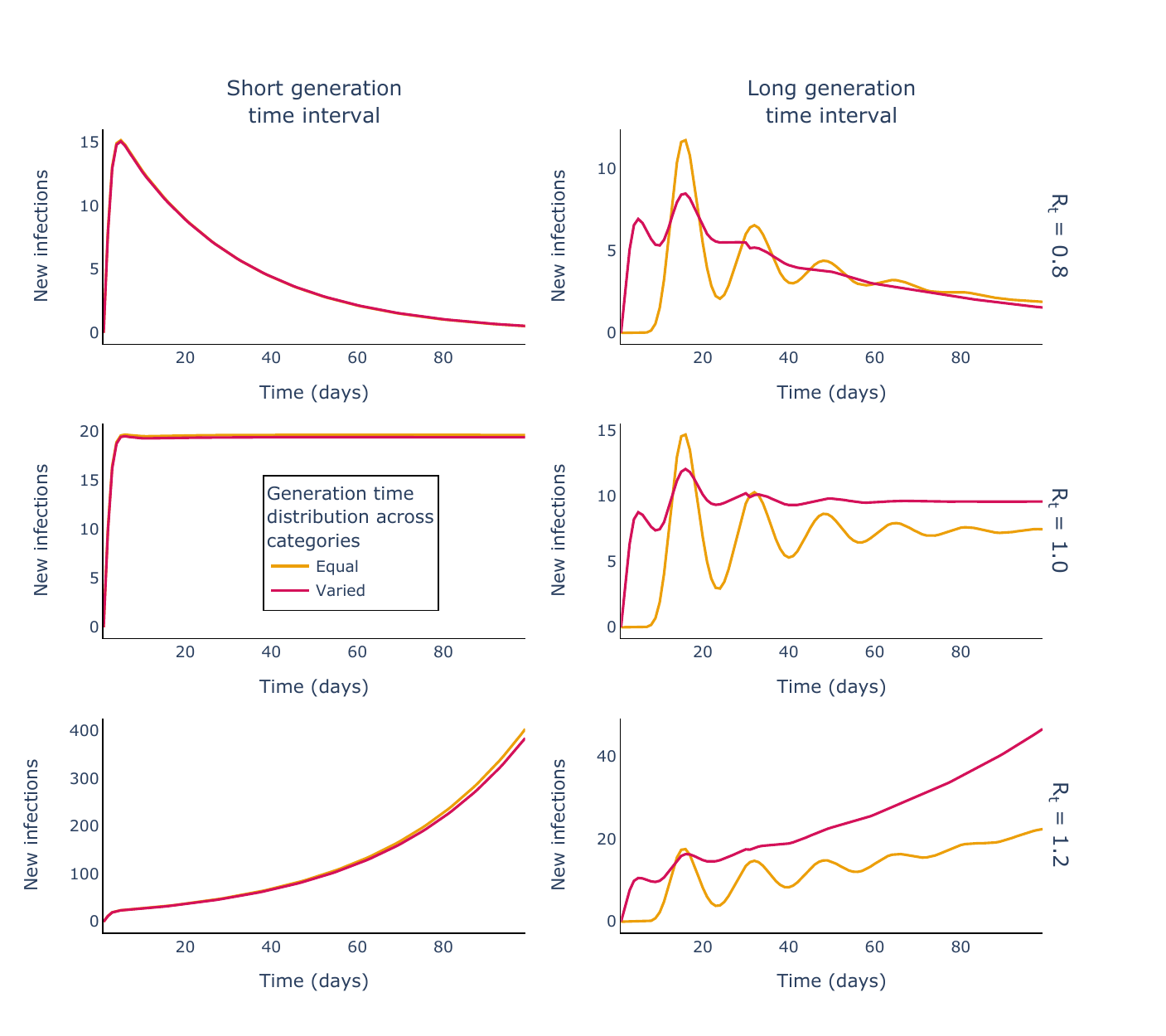}
\caption{\textbf{Whether an outbreak shrinks, stabilises or grows in the long-term does not depend on the generation time distribution.} The rows correspond to different $R_t$ values, increasing from top to bottom. The columns correspond to short (left) and longer (right) generation times. The yellow lines show aggregate outbreak dynamics (i.e. the evolution of the sum of new infection counts across the three groups) when the generation time distribution is the same across all groups (given by scenarios 5 and 2 in Table \ref{tab:age_group_generation_times} for the short and long generation time intervals, respectively); the red lines show the equivalent quantity when there is variation in the generation time distributions across the groups (the assumed generation times correspond to scenario 4 for the short generation time intervals and scenario 1 for the long generation time intervals in Table \ref{tab:age_group_generation_times}).}
\label{Different SI - Growth rate}
\end{figure}
We now use simulations from eq. \eqref{eq:discrete_renewal_multigroup} to illustrate this. We again assume that the population is structured into three age groups as described in \S\ref{sec:time_varying_structured_Rt}.

In Figure \ref{Different SI - Growth rate}, we show the results of our simulations of unfolding outbreaks. This first set of simulations is based on two assumptions around variation in the generation time distribution across the three groups: i) when the three groups have the same generation time distribution (red lines), or ii) each of the groups have different generation time distributions (yellow lines). We explore how the mean of the generation time distribution affects outbreak trajectories: the left-column shows simulations when the mean generation time interval is typically short; the right-hand column shows simulations when the generation time interval is generally longer. Descending the rows corresponds to increasing $R_t$ values.

Figure \ref{Different SI - Growth rate} shows that the $R_t$ threshold determines whether, ultimately, an outbreak shrinks (top row: $R_t=0.8$), stabilises (middle row: $R_t=1$) or grows (bottom row: $R_t=1.2$). The assumed form of the generation time distribution across the groups does not affect whether longer-term infection counts decline, stabilise or grow. However, changes to the generation time distribution do affect the rate in calendar time at which infection counts change in the long term, and this is supported by further simulations (see \S\ref{sec:calendar_time}).

For short generation time intervals, the infection counts reach their smooth long-term trajectories rapidly after an initial period where they respond to the initial conditions. For long generation times, particularly when the generation time distribution varies by group, there are substantial oscillations in the infection counts, which appear to lessen with time. Further work is needed to understand these short-term behaviours of epidemic dynamics.

\subsection{Stochastic models have the same $R_t$ threshold governing their mean behaviour}

We now consider stochastic renewal models. In \S\ref{sec:stochastic_renewal_proof} we show that the reproduction number in a stochastic, discrete, single-population renewal model governs the long-term fate of the mean of the stochastic process; this proof carries directly over to the multiple-group case. We now illustrate via simulation that the expression derived for $R_t$ for the continuous model case (eq. \eqref{eq:rt_threshold}) governs the behaviour of the mean of a stochastic renewal process of the form:
\begin{align*}
     I_t^{(j)} &\sim \text{Poisson}\Big(\frac{R_t}{\rho(C_t)} \sum^{N}_{i=1}C_t^{(ji)}\sum^{t-1}_{a=1}w_a^{(i)} I_{t-a}^{(i)}\Big),
\end{align*}
where $I_t^{(j)}$ denotes the incidence in group $j$. The dynamics of the aggregate incidence across all groups is then given by:
\begin{align}
I_t \sim \text{Poisson}\Big(\frac{R_t}{\rho(C_t)} \sum^{N}_{j=1} \sum^{N}_{i=1}C_t^{(ji)}\sum^{t-1}_{a=1}w_a^{(i)} I_{t-a}^{(i)}\Big).
\label{Multicategorical Final Form}
\end{align}
We perform simulations of eq. \ref{Multicategorical Final Form} and its deterministic counterpart in a three-group population assuming a contact matrix of the form given by eq. \eqref{eq:contact_matrix}.

\begin{figure}[h]
\centering
\includegraphics[width=1\linewidth]{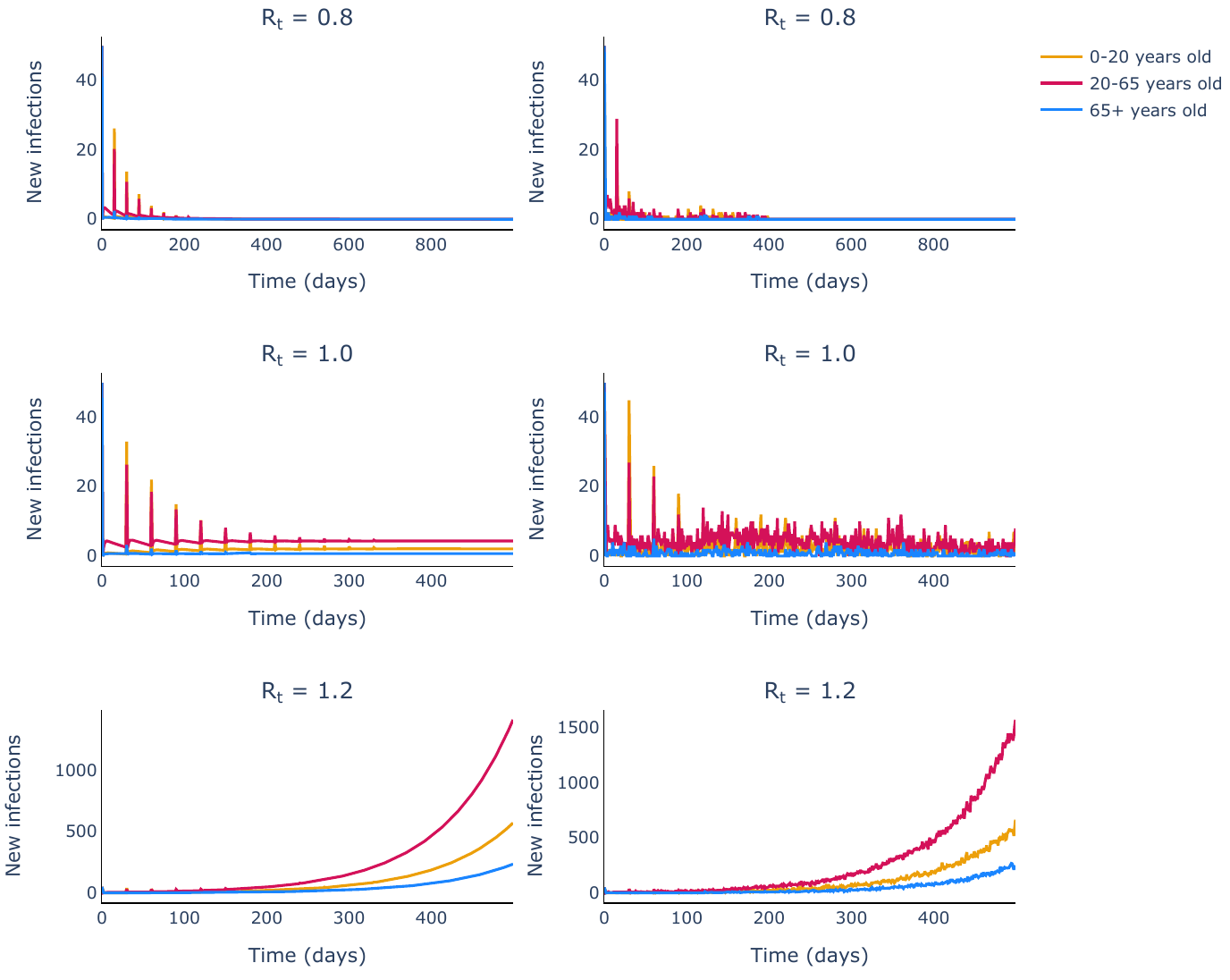}
\caption{\textbf{The long-term qualitative behaviour of incidence is determined in stochastic multiple-group renewal models by the same $R_t$ expression as for deterministic models.} The left-hand column shows simulations from a deterministic model and the right-hand column shows simulations from the stochastic model. The rows dictate the $R_t$ values assumed. For all the populations modelled we used the same age-specific generation time intervals as given by scenario 6 in Table \ref{tab:age_group_generation_times}.}
\label{Epidemic Fate}
\end{figure}

In Figure \ref{Epidemic Fate} we show that $R_t=\gamma_t \rho(C_t)=1$ is the boundary that dictates the fate of the aggregate incidence for the deterministic model (left-hand column) and, analogously, for the long-run mean aggregate incidence of the stochastic process.

\subsection{The relationship between $R_t$ and growth rate $r$ for structured populations}
\label{R vs r}
Considering a homogeneous population, if eq. \eqref{eq:euler_lotka} is divided through by $R_t=C(t)\gamma(t)$, we obtain:
\begin{equation}\label{One-cat growth rate}
    \frac{1}{R_t} = \int_0^t w(a) \exp(-r a) da.
\end{equation}
which previous studies have shown represents the relationship between $R_t$, the growth rate per infected individual and $r$, the growth rate of the epidemic size in calendar time \cite{Wallinga(2007)}.
We now derive an analogous relationship for a structured population. To do so, we take eq. \eqref{Phi multiple group} and divide this through by $R_t=\gamma_t \rho(C_t)$, we obtain:
\begin{align*}
    \frac{1}{R_t} &= \frac{ \rho(\overline{K}(r))}{\gamma_t \rho(C_t)} = \frac{\gamma_t \rho\Big(C_t\int_0^\infty W(a) \exp(-r a) da\Big)}{\gamma_t \rho(C_t)} \\ 
    \frac{1}{R_t} &= \frac{\rho\Big(C_t\int_0^\infty W(a) \exp(-r a) da\Big)}{\rho(C_t)}. \numberthis \label{eq:general_growth_rt}
\end{align*}
This shows that the relationship between $R_t$ and $r$ depends on the contact patterns in the population and the individual group generation times distributions. Interestingly, it does not depend on $\gamma_t$, the probability that a contact results in an infection (since we have modelled this as not varying by group).

An analogous relationship holds between $R_t$ and the doubling time of the epidemic (see \S\ref{sec:doubling_time}).

We now demonstrate that the relationship between $R_t$ and $r$ holds using simulations of an equivalent discrete-time model, which has a corresponding relationship between these two quantities:
\begin{equation}\label{eq:Rt_growth}
\frac{1}{R_t} = \frac{\rho\Big(C_t\sum_{a=0}^\infty W_a (1+r)^a\Big)}{\rho(C_t)},
\end{equation}
where $W_a$ is the diagonal matrix of the discretised generation time intervals evaluated at the discrete time interval $a$.

\begin{figure}[h]
\centering
\includegraphics[width=0.8\linewidth]{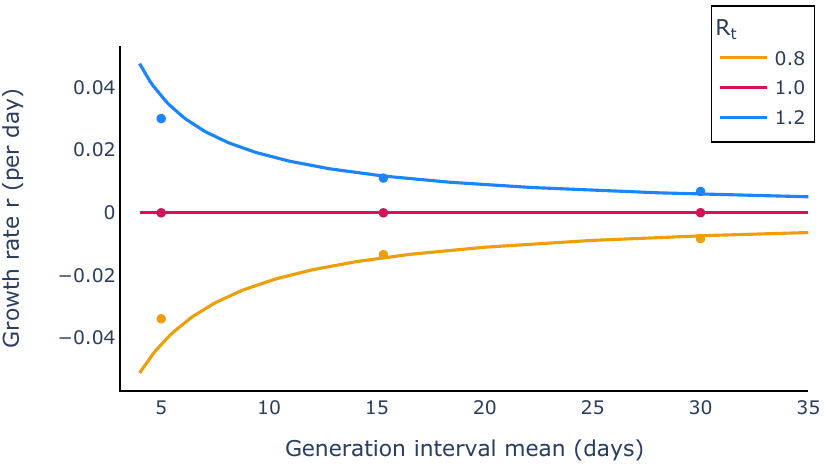}
\caption{\textbf{Theoretical and simulation-based relationships between epidemic growth and generation time means largely agree.} We represent in different colours the three possible epidemic long-term behaviours: epidemic growth ($R_t=1.2$, blue), epidemic persistence ($R_t=1$, red) and decay, respectively ($R_t=0.8$, yellow). The solid lines indicate the theoretical trajectories given by eq. \eqref{eq:Rt_growth}, and the points represent analogous estimates derived from simulations of the deterministic model, computed as seen in \S\ref{sec:empirical estimates of the growth rate from simulations}.}
\label{R-curves}
\end{figure}

In Figure \ref{R-curves} we show the relationship between $r$ and the mean of the generation time across three values of $R_t$. The lines show the theoretical quantities predicted by eq. \eqref{eq:Rt_growth}; the points show the empirical results derived from simulations. This plot shows that the theoretical and empirical results are in reasonable agreement. It also shows that as the generation time mean increases, the rate of change of epidemic size in calendar time declines (for those cases when $R_t\neq 1$).

\paragraph{Same generation time distributions across groups.} If the generation time distribution across all groups is the same, we can write $W(a)=w(a) \mathbb{I}_N$ and then use this to simplify eq. \eqref{eq:general_growth_rt}:
\begin{align*}
    \frac{1}{R_t} &= \frac{\rho\Big(C_t\int_0^\infty w(a) \mathbb{I}_N e^{-r a} da\Big)}{\rho(C_t)}= \frac{\rho\Big(C_t(\int_0^\infty w(a) e^{-r a} da)\Big)}{\rho(C_t)}\\
    &= \frac{\big(\int_0^\infty w(a) e^{-r a} da\big)\rho(C_t)}{\rho(C_t)}= \int_0^\infty w(a) e^{-r a} da, 
\end{align*}
to recover eq. \eqref{One-cat growth rate}. So, when all groups share the same generation time distribution, contact patterns do not affect the relationship between $R_t$ and $r$.

\paragraph{Varying generation time distributions across groups.} We now explore how the relationship between $r$ and $R_t$ is affected by the composition of generation time distributions across the groups.

We consider three scenarios making different assumptions about the generation time distributions in each age group. For each scenario and for each group, we assume the generation time distribution is governed by a (discretised) gamma distribution with the same standard deviation, of 3.3 days; the mean generation time is allowed to vary by group. In the first scenario, we assume the mean generation time is lowest in the youngest age group and highest in the eldest; in the second, the mean generation time is the same in each group; in the third scenario, the generation time decreases with age. The means in each scenario are given in Table \ref{tab:age_group_generation_times}.

\begin{figure}[h]
\centering
\includegraphics[width=0.9\linewidth]{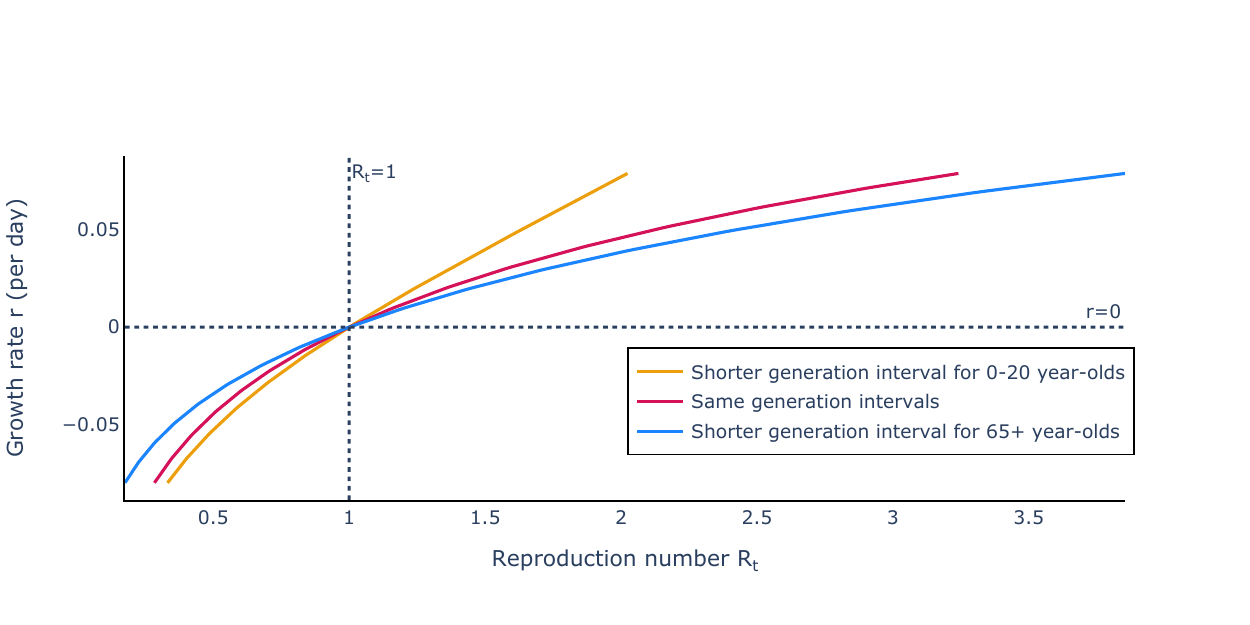}
\caption{\textbf{The relationship between $r$ and $R_t$ is shaped by subpopulation generation time distributions.} The yellow line corresponds to a population with a shorter generation time interval in the young population, equivalent to scenario 1. The red line corresponds to a simulation for which the generation time interval is identical in all population groups, with a mean of 15.3 as seen in scenario 2. Lastly, the blue line corresponds to a population with a shorter generation time interval in the old, equivalent to scenario 3.}
\label{fig:R vs r trajectory}
\end{figure}

In Figure \ref{fig:R vs r trajectory} we show the relationship between epidemic growth rate $r$ and $R_t$ implied by eq. \eqref{eq:general_growth_rt} according to each of the three scenarios. In all cases, $r=0$ implies $R_t=1$. Since the 0-20 age group has the highest contact rates, most infections are generated from this group and it has the single greatest contribution to $R_t$. Because of this, changes in the generation time of this group skew the relationship between $r$ and $R_t$: when the generation time interval is short in this group, the corresponding growth rate is high; and this decreases as their generation time is increased.

\begin{figure}[h]
\centering
\includegraphics[width=1\linewidth]{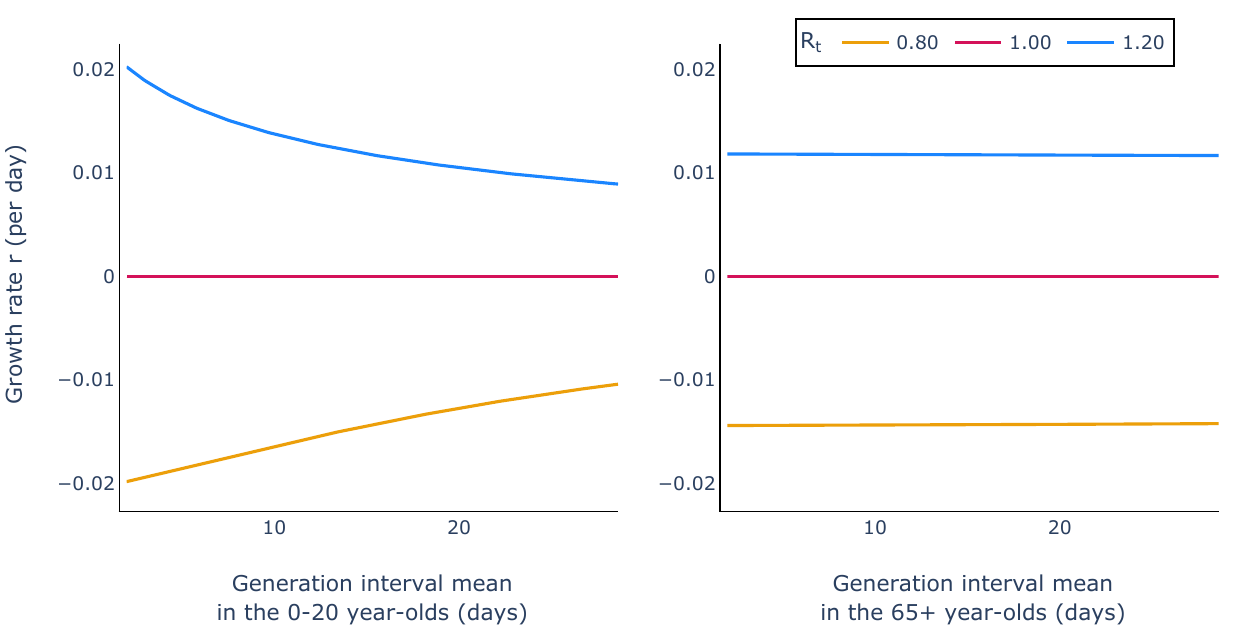}
\caption{\textbf{The epidemic growth rate is sensitive to changes in the mean generation time in high contact groups.} In both panels we show the relationship between the epidemic growth rate in calendar time and the mean generation time interval in a particular group: left, those aged 0-20; right, those aged over 65. When varying the generation time mean in one group (between $2$ and $28.6$ days, we hold the means in the other groups constant at $15.3$ days. For all age groups, we considered an equal standard deviation of $3.3$.}
\label{fig:Shorter generation intervals}
\end{figure}

To further highlight this sensitivity, in Figure \ref{fig:Shorter generation intervals} we show how the epidemic growth rate responds to changes in the mean generation time in those aged 0-20 (left-hand panel) and those aged 65+ (right-hand panel); in each case, the mean generation time in the other groups remains fixed at 15.3 days. We consider three values of $R_t$ for each panel (corresponding to the coloured lines). As intuited above, the growth rate is more sensitive to variations in the mean generation time for the higher contact 0-20 age group.

\paragraph{Homogeneous mixing} We now suppose that all groups intermix at the same rate and derive a relationship between $R_t$ and $r$ in this limit. Assuming groups mix homogeneously amounts to assuming an $N\times N$ contact matrix of the form:
$$
C_t=c\begin{bmatrix}
    1 & \dots & 1 \\
    \vdots & \ddots & \vdots\\
    1& \dots & 1
  \end{bmatrix}.
$$
In this case, $\rho(C_t)=c N$ (see \S\ref{Identical rows matrix spectral radius}) meaning the reproduction number $R_t=cN\gamma_t$. We now consider:
$$C_t\int_0^\infty W(a) e^{-r a} da = \begin{bmatrix}
    c\int_0^\infty w^1(a) e^{-r a} da & \dots & c\int_0^\infty w^N(a) e^{-r a} da \\
    \vdots & \ddots & \vdots\\
    c\int_0^\infty w^1(a) e^{-r a} da& \dots & c\int_0^\infty w^N(a) e^{-r a} da
  \end{bmatrix},$$

which has spectral radius $\rho\Big(C_t\int_0^\infty W(a) e^{-r a} da\Big) = c\int_0^\infty \sum_{i=1}^N w^i(a) e^{-r a} da$ (see \S\ref{Identical rows matrix spectral radius}). This suggests the following relationship between the reproduction number and growth rate:
\begin{align*}
    \frac{1}{R_t} = \frac{c\int_0^\infty \sum_{i=1}^N w^i_s e^{-r a} da}{cN} = \int_0^\infty \Big(\frac{1}{N}\sum_{i=1}^N w^i(a)\Big) e^{-r a} da,
\end{align*}
which is equivalent to that of a single population with a generation time distribution with elements $\bar w(a) := \frac{1}{N}\sum_{i=1}^N w^i(a)$; in other words, with a generation time distribution equal to the average across all groups.

\section{Discussion}
The time-dependent reproduction number, $R_t$, is a critical measure for tracking pathogen transmission during an infectious disease epidemic. Modelling frameworks for inferring $R_t$ have tended to neglect heterogeneities between host groups in the population under consideration. However, heterogeneous mixing patterns and differences in the susceptibility and progression of a disease across groups can play an important role in shaping the epidemic trajectory. Here, we have developed a renewal equation framework enabling $R_t$ to be determined while accounting for these heterogeneities.

Our analyses produced an expression for $R_t$ which, as expected, is not contingent on the generation time distribution; we show that our $R_t$ expression determines the long-term qualitative dynamics of the epidemic (i.e., whether case numbers eventually grow or decline). This is despite the fact that the generation time sets the relationship between $R_t$ and the epidemic growth rate (Figure \ref{R-curves}). Finally, we showed that our $R_t$ expression also governs the expected long-term dynamics of stochastic renewal equation models.

For the stochastic renewal equation underlying parts of our analysis, we assumed that the number of cases each day is drawn from a Poisson distribution. However, more over-dispersed distributions can be used in renewal equation models, including accounting for the possibility of super-spreading events \cite{ho2023accounting,Thompson(2024)}. So long as the mean of the process is the same as we assumed, however, our results should hold.

The research presented here builds on a substantial amount of literature on the topic of $R_t$ inference. Initial methods for estimating $R_t$ have been extended in multiple directions, including allowing for different strains of a pathogen \cite{bhatia2023extending} and inference of the generation time or serial interval distribution \cite{Thompson(2019)}. In recent work, temporal aggregation of disease incidence data (e.g., weekly rather than daily case reporting) has been accounted for in $R_t$ inference methods \cite{nash2023estimating,ogi2024simulation}. Despite these extensions, an exhaustive analysis of $R_t$ quantification in populations consisting of multiple host groups has not been undertaken until now.

As with any modelling study, the approach presented here relies on various assumptions. For robust quantification of $R_t$ using renewal equations, reliable estimates of the generation time distribution are required. Such estimates should account for the possibility that this quantity can change during an ongoing outbreak \cite{Chen(2022),hart2022inference,hart2022generation,Parag(2023)b}. We additionally assumed that all infections are generated within the populations modelled; that is, there are no imported infections, and failing to account for their presence can lead to biases in inferred $R_t$ values \cite{Creswell(2022)}. An $R_t$ inference method based on our work here would also require estimates of the contact matrix and the relative infectiousness and susceptibility for those groups within the population under consideration; information on group-specific case counts would also provide useful information for $R_t$ quantification. Additionally, the renewal equation model that we derived is only suitable for modelling directly transmitted pathogens. Extending this work to model heterogeneities that are relevant for other pathogens, such as vector-borne diseases, is another possible target for future research. 

Despite these simplifications, our work provides a theoretical basis for future studies that account for heterogeneity in host populations when inferring $R_t$. Using our analytically derived expression for $R_t$, the effects of a range of interventions on the epidemic fate could be explored. Non-pharmaceutical interventions (NPIs) may impact population mixing patterns, which would change the effective contact matrix $C_t$ and, consequently, change $R_t$. This means that we can assess the impact of some NPIs on $R_t$ by simply changing the contact matrix structure. As an example, the impacts of interventions that affect particular age groups, such as school or workplace closures, could be explored by modifying the contact matrix \cite{lovell2022estimating}. Undertaking such an analysis in a realistic fashion would require real-time estimates of contact matrices, which could be provided from, for example, mobility data \cite{di2024mobility}.

Renewal equation models have proved remarkable tools in epidemic surveillance, owing to the relatively few assumptions required to infer $R_t$. An implicit assumption underpinning the majority of their usage is, however, that there are no important heterogeneities in the population. Here, we show how to account for these heterogeneities in a renewal equation framework for structured populations. With the advent of real-time data on individual contact patterns, these modified renewal equation approaches could form the basis of future methods for inferring epidemiological quantities such as $R_t$. By enabling public health policy-makers to track transmission more accurately, such approaches have the potential to allow policy-makers to make the most informed possible decisions during future outbreaks of a wide range of pathogens.


\section{Acknowledgements}

IB would like to thank Paveluta Bouros for her help in identifying the applicability of Thieme's theorem to this work.

IB and RNT would like to thank members of the Infectious Disease Modelling group in the Mathematical Institute at the University of Oxford for useful discussions about this work.

\beginsupplement

\makeatletter
\renewcommand \thesection{S\@arabic\c@section}
\renewcommand\thetable{S\@arabic\c@table}
\renewcommand \thefigure{S\@arabic\c@figure}
\makeatother

\appendix
\setcounter{section}{0}

\section{Appendix}

\subsection{Thieme's renewal theorem}\label{sec:thieme_renewal}
Here, we closely follow the more general approach presented in \cite{Boldin(2023)} for a continuous-time renewal equation model with multiple groups.

We reframe our modelling framework as follows:

\begin{align*}
    \begin{pmatrix}
I^1(t) \\
\vdots \\
I^N(t) 
\end{pmatrix} = \int_{a=0}^\infty K(a) \begin{pmatrix}
I^1(t-a) \\
\vdots \\
I^N(t-a) 
\end{pmatrix} da,
\end{align*}

\noindent where the kernel of the renewal process is given by $K(a)= \gamma_t C_t W(a)$. If the kernel satisfies a set of conditions (which we give below), then the long-term solution can be shown to follow:
\begin{align*}
    \begin{pmatrix}
I^1_t \\
\vdots \\
I^N_t 
\end{pmatrix} = e^{r t} \underline{\Phi}.
\end{align*}

The kernel function must then satisfy the following five conditions:
\begin{enumerate}
    \item $K(a)$ is a non-negative matrix for all $a>0$. This is true, since $\gamma_t  > 0$, and $W(a)$ and $C_t$ are non-negative matrices.
    \item An element of $K(a)$ exceeds zero for some $a>0$. Similar to before, this is also satisfied as $\int_{a=0}^\infty K(a)da = \gamma_t  C_t \int_{a=0}^\infty W(a)da = \gamma_t  C_t$ has at least two non-zero elements (otherwise all contacts are $0$).
    \item $\int_{a=0}^\infty \norm{K(a)} da < \infty$, where we define the $L_1$-norm of a matrix $M$ as $\norm{M}=\max_{1 \leq i \leq N}\sum_{j=1}^N \abs{M^{(ji)}}$ and $\abs{M^{(ji)}}$ is the absolute value of the $(j, i)^\text{th}$ element of matrix $M$.

\noindent This condition is also met: by replacing the kernel function evaluated at $a$ with $\gamma_t  C_t W(a)$, the conditions becomes

\begin{align*}
    \int_{a=0}^\infty \norm{K(a)} da &= \gamma_t  \int_{a=0}^\infty \max_{1 \leq i \leq N}\sum_{j=1}^N \abs{C_t^{(ji)}w^i(a)} da= \\
    & = \gamma_t  \Big( \int_{a=0}^\infty w^i(a) da \Big) \max_{1 \leq i \leq N}\sum_{j=1}^N \abs{C_t^{(ji)}} = \\
    & = \gamma_t  \max_{1 \leq i \leq N}\sum_{j=1}^N C_t^{(ji)} < \infty
\end{align*}

\noindent as required. For the last two conditions, Boldin et al \cite{Boldin(2023)} uses a discrete time renewal process. We follow with similar assumptions which we assume to hold in the continuous time case as well.

    \item There exists $a_0 \in \mathbb{N}$ such that $R(a_0)$ is a strictly positive matrix, where $R(a) = R * K(a) + K(a) = K * R(a) + K(a), \forall a>0$ is defined as the resolvent of the kernel matrix $K(a)$. We use $R * K(a)$ to mean $R * K (a) = \sum_{t=0}^a R(a-t)K(t)$ as described in \cite{Thieme(1984)}; also a strictly positive matrix implies that all its elements are greater than $0$; that is $m_{ij}>0, \forall i, j$ for a strictly positive matrix $M=(m_{ij})_{i,j}$.
    \item Additionally, there exists $a_1\in \mathbb{N}$ and $a_2\in \mathbb{N}$ with greatest common divisor $(a_1, a_2)=1$ such that both $R(a_1)$ and $R(a_2)$ have all elements greater than $0$ and at least one element exceeding $0$.
\end{enumerate}

For the fifth condition, we recursively compute the values of resolvent matrix $R(a)$. For $a=0$, we have that $R(0)=K(0)=\mathbb{O}_N$ as seen in \cite{Boldin(2023)}. For the next value $a=1$, we have that $R(1)=K(1)+W(0)R(1)+W(1)R(0)=K(1)$. As $K(1)=\gamma_t C_t W(1)$ is a non-negative matrix, then $R(1)$ is a non-negative matrix.
From the definition of $R(a)$, by induction we conclude that $R(a)$ is a non-negative matrix for all $a \in \mathbb{N}$. We consider the first time point $a'$ for which $K(a')$ has an element bigger than zero. From the second condition we have that such an $a'$ exists. 

If $a'=1$, we have that both $R(2)=K(2)+K(1)R(1)=K(2)+K(1)^2$ and $R(3) = K(3) + K(1)R(2)+K(2)R(1) = K(3) + K(1)K(2) + K(2)K(1) + K(1)^3$ will have at least one element greater than zero and are non-negative matrices, and as $(2,3)=1$, the desired fifth condition is satisfied.

If $a' \geq 2$, however, $R(a') = K(a') + R *K(a')$ has an element bigger than zero as $R*K(a')$ is itself a non-negative matrix. Additionally, $R(a'+1) = K(a'+1) + K(a')R(1) + \dots K(1)R(a') = K(a'+1) + K(a')K(1) = K(a'+1)$ as the condition on the value of $a'$ implies that $K(a)$ is the identical zero matrix $\forall a < a'$ and $R(1)=K(1)$.

In terms of the fourth condition, we saw from before that $R(a)$ can be written an a sum of powers of the kernel matrices $K(b)$ for some $b$s with $b \leq a$ for all $a \in \mathbb{N}$. The kernel matrices $K(a)= \gamma_t C_t W(a)$, where $W(a)$ is a diagonal matrix. If all population groups mix between each other, the contact matrix $C_t$ will be strictly positive. Therefore, $K(a')$ will also be strictly positive, satisfying the fourth condition.

Otherwise, if the modelled population includes demographic groups that do not interact with any other groups, then the contact matrix $C_t$ is a diagonal matrix and can be written as 

\begin{equation}
\label{non-mixing Ct}
C_t = 
\begin{pNiceArray}{ccc|ccc}
  c_{11} & & & \Block{3-3}<\Large>{\mathbf{0}} \\
  & \ddots & & & &\\
  & & c_{mm} & & &\\
  \hline
  \Block{3-3}<\Large>{\mathbf{0}} & & &\\
  & & & & \Large C^{*}_t &\\
  & & & & &\\
\end{pNiceArray}.
\end{equation}

As $W(a)$ is a diagonal matrix itself, then we would have that the kernel matrix evaluated for any time $a$ is equal to
\begin{equation*}
    K(a) = \gamma_t
    \begin{pNiceArray}{ccc|ccc}
  c_{11}w^1(a) & & & \Block{3-3}<\Large>{\mathbf{0}} \\
  & \ddots & & & &\\
  & & c_{mm}w^m(a) & & &\\
  \hline
  \Block{3-3}<\Large>{\mathbf{0}} && & & &\\
  & & & & C^{*}_t W_{m:, m:}(a) &\\
  & & & & & \\
    \end{pNiceArray}.
\end{equation*}

Regardless of the number of how many times this type of matrix is raised to any power and how many times we sum it to compute $R(a)$, the top block of the resulting matrix will remain a diagonal matrix and therefore not satisfying the forth condition imposed. Hence, our findings are confined to only fully mixing populations.

\subsection{Renewal equations and reproduction numbers for a discrete-time structured population model}\label{sec:discrete_renewal_theory}
When inferring $R_t$, the renewal equation frameworks typically used are discrete in time (usually with time steps of 1 day; see \S\ref{Methods}). We now demonstrate that for a discrete-time version of the model presented so far, we obtain an analogous renewal equation as for the continuous-time model and the same expression for $R_t$.

When time and age are discrete, the analogous conservation equation for the population is a difference equation:
\begin{align*}
    \frac{n(t+\Delta t, a+ \Delta a)- n(t, a+ \Delta a)}{\Delta t}+ \frac{n(t, a+ \Delta a)- n(t, a)}{\Delta a} = 0,
\end{align*}
where $\Delta t = \Delta a$ represents a discrete unit of time. The density of new infections occurring at a time $t=T\Delta t$ is:
\begin{equation} \label{eq:nearly_renewal discrete}
    n(T \Delta t,0) = \sum_{A=0}^\infty C_t \gamma_t w(A\Delta t) n(T \Delta t, A \Delta t).
\end{equation}
Similar to the continuous-time population model, we only consider infections occurring after time $t=T\Delta t=0$. The initial number of infections at time $T=0$ can be written as $n(0, A\Delta t)= f(A\Delta t)$. The analogous characteristics for this discrete-time model are:
\begin{equation}\label{eq:characteristics_discrete}
a= A \Delta t = \begin{cases}
T \Delta t + A_0 \Delta t,\text{ for }a > t\text{ i.e.\ } A>T\\
T \Delta t - T_0 \Delta t,\text{ for }a \leq t\text{ i.e.\ } A\leq T.
\end{cases}
\end{equation}
\noindent where $A_0 >0$ is the number of age units of an infection at time $t=0$ and $T_0 \geq 0$ is the number of units of time at which an infection occurred. This allows us to define the number of infections as a function of $t$ only, that is we can write $n(T\Delta t, A\Delta t) = n(T\Delta t, A(T)\Delta t) := n(T\Delta t)$, for a given choice of $A_0$, or $T_0$ respectively. Along each such defined \textit{characteristic} lines, the number of infections is determined by a singular point in time $\tau$, with $\tau=A_0\Delta t$ uniquely for $A>T$ and $\tau=T_0\Delta t$ for $A \leq T$. Therefore, the number of infections is conserved along each characteristic line, i.e.\
\begin{equation*}
    n(t,a(t)) = \text{const.}
\end{equation*}
Using eqs. \ref{eq:characteristics_discrete}, we derive the constant population sizes: when $A>T$, $A$ can be written $a=A \Delta t= T\Delta t + A_0 \Delta t$ which implies $n(T \Delta t,T \Delta t+A_0 \Delta t)=\text{const.}$; similarly when $T=0$, the population size is:
\begin{equation}\label{eq:a_above_t discrete}
    n(0, A_0 \Delta t)=f(A_0 \Delta t)=f((A-T)\Delta t).
\end{equation}
When $A\leq T$, we use instead the fact that $A \Delta t=T \Delta t-T_0\Delta t$; as $n(t,t-T_0\Delta t)=\text{const.}$, and, when $T=T_0$:
\begin{equation}\label{eq:t_above_a discrete}
    n(T_0\Delta t,0)=n((T-A)\Delta t,0).
\end{equation}
Collecting eqs. \eqref{eq:a_above_t discrete} \& \eqref{eq:t_above_a discrete}, the solutions for the discrete population sizes for all infection onset times are given by either of the following two forms:
\begin{equation}\label{eq:system_solutions discrete}
    n(T \Delta t, A \Delta t)= \begin{cases}
    f((A-T)\Delta t), A>T,\\
    n((T-A)\Delta t,0), A\leq T.
\end{cases}
\end{equation}
Substituting the eqs. \ref{eq:system_solutions discrete} back into \ref{eq:nearly_renewal discrete}, we obtain that
\begin{align*}
    n(T\Delta t, 0) &= \sum_{A=0}^T b(A\Delta t) n(T\Delta t, A \Delta t)+ \sum_{A=T+1}^\infty b(A\Delta t) n(T\Delta t, A \Delta t)=\\
    &=\sum_{A=0}^T b(A\Delta t) n((T-A)\Delta t, 0) + \sum_{A=T+1}^\infty b(A\Delta t) f((A-T)\Delta t).
\end{align*}
Since similarly to the continuous-time case,  for any $A'>T$, $b(A\Delta t) \approx 0$, the number of new infections observed at time $t=T \Delta t$ becomes:
\begin{align*}
    n(T\Delta t, 0)&= \sum_{A=0}^T b(A\Delta t)n((T-A)\Delta t, 0) + \sum_{A=T+1}^\infty 0 f((A-T)\Delta t)=\\
    &= \sum_{a=0}^t \gamma_t  C_t w_a n(t-a, 0) = \sum_{a=0}^t \gamma_t  C_t w_a I_{t-a},
\end{align*}
\noindent which is indeed identical to the deterministic form of the renewal equation process as described in \eqref{Uni-categorical Process}. The rest of the proof for the definition of the reproduction number and epidemic stability criterion $R_t = \sum_{a=0}^\infty \gamma_t  C_t w_a = \gamma_t  C_t \sum_{a=0}^\infty w_a = \gamma_t  C_t$ follows according to the methodology presented in \cite{Boldin(2023)}.

For the multiple group population, a similar argument emerges \cite{Boldin(2023)} -- we arrive at the conclusion that the overall reproduction number $R_t = \rho(\sum_{a=0}^\infty \gamma_t  C_t W_a) = \rho(\gamma_t  C_t \sum_{a=0}^\infty W_a) = \rho(\gamma_t  C_t) = \gamma_t\rho(C_t)$ is a correct criterion for the long-term behaviour of an epidemic.

\subsection{The typical generation time interval affects epidemic growth rates in calendar time}\label{sec:calendar_time}

\begin{figure}[H]
\centering
\includegraphics[width=0.9\linewidth]{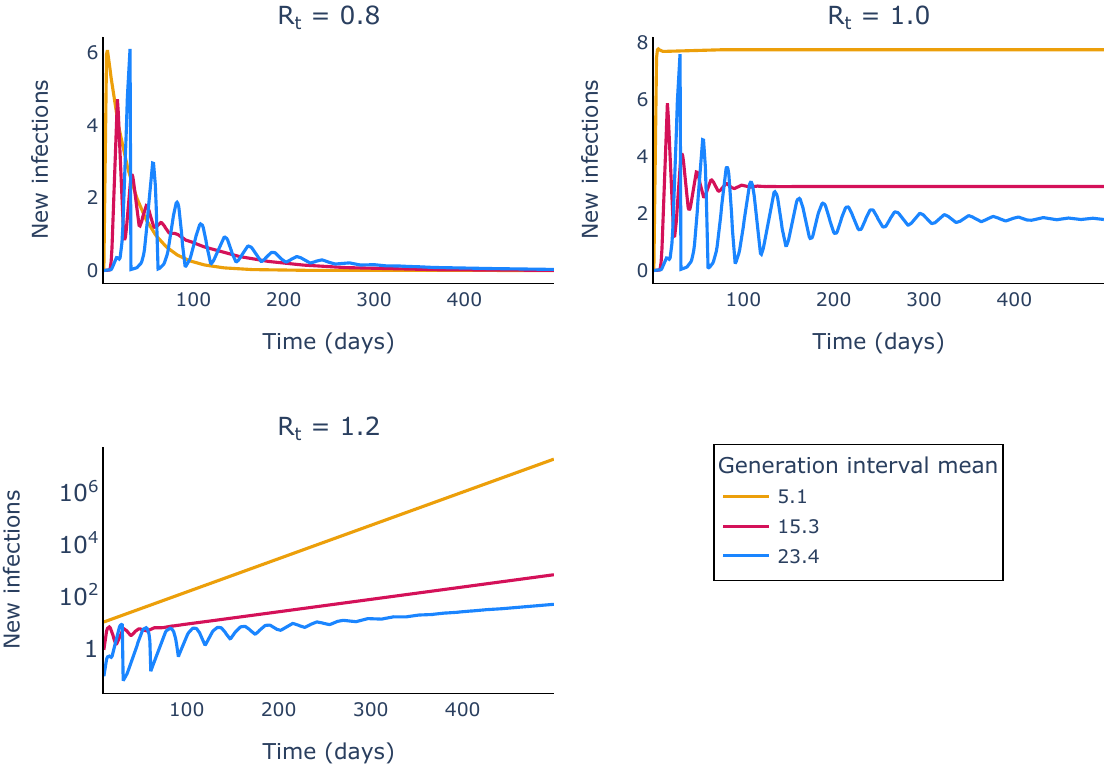}
\caption{\textbf{The typical generation time interval affects epidemic growth rates in calendar time.} Each panel represents a different $R_t$ value, as indicated. Each line represents aggregate new infection counts (i.e. the aggregate across three populations) for a different simulation scenario: the line colours represent these scenarios and correspond to differing average generation time distributions across the three groups as indicated in the legend.}
\label{Different SI}
\end{figure}

In Figure \ref{Different SI}, we simulate the trajectories of the total number of new infections for a range of values of the generation time intervals means:

$$\overline{W}_1 = \begin{pmatrix}
5 \\
5.3 \\
5
\end{pmatrix}, \overline{W}_2 = \begin{pmatrix}
15.3 \\
15.3 \\
15.3
\end{pmatrix} \text{ and } \overline{W}_3 = \begin{pmatrix}
30 \\
25.3 \\
15
\end{pmatrix}.$$

In each case, the standard deviation of the generation time distribution was assumed to be 3.3 days.

\noindent We use the contact matrix given by eq. \eqref{eq:contact_matrix}. The infection rate $\gamma_t$ is also kept constant for all plotted scenarios. In the beginning of the simulation we see the different oscillatory patterns for each of our choices of the generation time intervals. However, as the time increases, for all $R_t$ values and all selected $\overline{W}$s, the oscillations die out and the number of new infections approach approach $0$ (panel A; epidemic decay), reach stable values (panel B; epidemic persistence) or continue to increase exponentially at a constant rate (panel C; epidemic growth).

\subsection{$R_t$ for a discrete-time stochastic model}\label{sec:stochastic_renewal_proof}

We now derive $R_t$ for a stochastic version of a renewal model. We consider only a renewal equation for a population comprising members of a single group since these results directly translate to the multiple group case. The renewal equation we assume is of the form:
\begin{equation}
I_t \sim \text{Poisson} \left( R \sum_{s=1}^{p} \omega_s I_{t-s} \right).
\end{equation}

This means that for an arbitrary time $t+\tau$, we have:

\begin{equation}
I_{t+\tau} \sim \text{Poisson} \left( R \sum_{s=1}^{p} \omega_s I_{t+\tau-s} \right).
\end{equation}

Taking expectations at time $t$ yields:

\begin{equation}
\mathbb{E}_t \left[ I_{t+\tau} \right] = R \sum_{s=1}^{p} \omega_s \mathbb{E}_t \left[ I_{t+\tau-s} \right].
\end{equation}

Then denoting $M_j := \mathbb{E}_t \left[ I_{t+\tau} \right]$ as the expectation at time $t$ of the infection count at time $t+\tau$ yields a renewal equation in terms of this quantity:

\begin{equation}
M_j = R \sum_{s=1}^{p} \omega_s M_{j-s}.
\end{equation}

We then suppose a long-run solution of the form $M_j \approx \lambda^j$, which results in:

\begin{equation}
\lambda^j = R \sum_{s=1}^{p} \omega_s \lambda^{j-s} \Rightarrow 1 = R \sum_{s=1}^{p} \omega_s \lambda^{-s}.
\end{equation}

We can then define:

\begin{equation}
\label{k_lambda}
K(\lambda) = R \sum_{s=1}^{p} \omega_s \lambda^{-s}.
\end{equation}

By construction, \( K(1) = R \) and \( K(\lambda) \) is a decreasing function of \( \lambda \). This means that:

\begin{itemize}
    \item If \( R > 1 \), \( \lambda^* = \lambda \text{ s.t. } K(\lambda) = 1 \) must satisfy \( \lambda^* > 1 \),
    \item If \( R \leq 1 \), \( \lambda^* \leq 1 \),
\end{itemize}

and similarly so for complex \( \lambda \) (see \cite{Boldin(2023)} section 2.1). This means that we can consider $R$ to satisfy the threshold conditions governing the long-run fate of the stochastic mean in the stochastic renewal model.

\subsection{Doubling times}\label{sec:doubling_time}
We derive the relationship between the reproduction number and the doubling time $\tau_0:= \frac{1}{r}\ln(2)$ -- the time taken for an epidemic to double in size. Rearranging this expression and substituting it in place of $r$ in eq. \eqref{eq:general_growth_rt} results in

\begin{align*}
    \frac{1}{R_t} = \frac{\rho\Big(C_t\int_0^\infty W(a) 2^{-\tau_0^{-1} a} da\Big)}{\rho(C_t)}.
\end{align*}

\subsection{The relationship between $R_t$ and the growth rate for the discrete-time deterministic model}\label{sec:discrete_rt-to-growth_rate}

We derive the relationship between  $R_t$ and the growth rate $r$ for structured populations in the case of the discrete-time deterministic model. From \cite{Boldin(2023)} section 2.2 and appendix \label{sec:discrete_renewal_theory} the overall reproduction number of the system is defined by $R_t = \rho(K(1))$, where $K(\lambda)$ is defined as

\begin{align*}
    K(\lambda) = \gamma_t C_t \sum_{a=0}^\infty W(a) \lambda^{-a};
\end{align*}

\noindent this definition for $K(\lambda)$ is simply the multivariate equivalent that eq. \eqref{k_lambda}. Similar to appendix \ref{sec:stochastic_renewal_proof}, $\lambda$ is defined in terms of the long-term solution of the infection vector count, that is $ \underline{I}_{j} = \lambda^j \underline{\Phi}$, where $\underline{\Phi}$ is a constant non-negative vector for large values of $j$. For the discrete-time deterministic model, the overall growth rate $r$ is defined such that $\underline{I}_{t+1} = (1+r) \underline{I}_t$, therefore implying that $\lambda = 1+r$. Hence, we have the following system of identities:

\begin{align}
    1 &= \rho(K(\lambda)) = \gamma_t \rho\Big(C_t \sum_{a=0}^\infty W(a) \lambda^{-a} \Big) = \gamma_t \rho\Big(C_t \sum_{a=0}^\infty W(a) (1+r)^{-a} \Big)\\
    R_t &= \rho(K(1)) = \gamma_t \rho\Big(C_t \sum_{a=0}^\infty W(a) \Big) = \gamma_t \rho(C_t), 
\end{align}

\noindent which when divided produce eq. \ref{eq:Rt_growth}. 

\subsection{Empirical estimates of the growth rate from simulations of the stochastic model}\label{sec:empirical estimates of the growth rate from simulations}

In order to determine the empirical growth rate $r$ of a population for which we know its overall reproduction number $R_t$, we use the scipy.integrate and scipy.optimize.minimize functions \cite{Scipy}. The goal function for the optimiser is given by the following equation in terms of $r$ derived from eq. \eqref{eq:general_growth_rt}:

\begin{align*}
    f(r) := \abs{\frac{\rho(C_t)}{\rho\Big(C_t\int_0^\infty W(a) \exp(-r a) da\Big)} - R_t}.
\end{align*}

The optimiser function will then return the value of $r$ for which $f(r)$ is closest to zero and thus approximates the empirical growth rate of the population thus simulated.

\subsection{Identical rows matrix spectral radius}\label{Identical rows matrix spectral radius}

We are computing the spectral radius of a positive matrix with identical rows, which is defined as the maximum real eigenvalue of the matrix. That is, the maximum value of $\lambda$ where 

$$\begin{bmatrix}
    a_1 & \dots & a_N \\
    \vdots & \dots & \vdots\\
    a_1 & \dots & a_N
  \end{bmatrix}\begin{pmatrix}
x_1 \\
\vdots \\
x_N 
\end{pmatrix} = \begin{pmatrix}
a_1 x_1 + \dots a_N x_N \\
\dots \\
a_1 x_1 + \dots a_N x_N
\end{pmatrix} = \lambda \begin{pmatrix}
x_1 \\
\vdots \\
x_N 
\end{pmatrix}$$

\noindent which implies $a_1 x_1 + \dots a_N x_N = \lambda x_i, \forall i$. One solution is for $\lambda=0$, which implies that $a_1 x_1 + \dots a_N x_N = 0$. Otherwise, if $\lambda \neq 0$ then that implies that $x_1= \dots x_N$, therefore $a_1 x_1 + \dots a_N x_N = (\sum_{i=1}^N a_i) x_1 = \lambda x_1$. Since we cannot have $x_1=0$ (otherwise the eigenvector is the zero-vector, contradiction), then the only other eigenvalue of the matrix is given by $\lambda= \sum_{i=1}^N a_i > 0$ by the definition of the positive matrix. Therefore the spectral radius of the matrix

$$\rho\Big(\begin{bmatrix}
    a_1 & \dots & a_N \\
    \vdots & \dots & \vdots\\
    a_1 & \dots & a_N
  \end{bmatrix}\Big)=\sum_{i=1}^N a_i.$$

\subsection{Exact overall reproduction number for a two-group population}

For the more general behaviour of a two group population we use the property that the spectral radius of a general 2x2 matrix

$$
M=\begin{bmatrix}
    a & b \\
    c & d
  \end{bmatrix}
$$

\noindent is given by the maximum eigenvalue, i.e.\ $\max\Big(\frac{a+d \pm \sqrt{(a-d)^2+4bc}}{2}\Big) = \frac{a+d + \sqrt{(a-d)^2+4bc}}{2}$. Therefore, if the contact matrix of the population is given by

$$
C_t=\begin{bmatrix}
    c_{11} & c_{12} \\
    c_{21} & c_{22}
  \end{bmatrix}
$$

\noindent then we have that $\rho(C_t)=\frac{c_{11}+c_{22} + \sqrt{(c_{11}-c_{22})^2+4c_{12}c_{21}}}{2}$ and similarly, using the fact that

$$C_t\int_0^\infty w_a e^{-r a} da = \begin{bmatrix}
    c_{11}\int_0^\infty w^1_s e^{-r a} da  & c_{12}\int_0^\infty w^2_s e^{-r a} da \\
    c_{21}\int_0^\infty w^1_s e^{-r a} da & c_{22}\int_0^\infty w^2_s e^{-r a} da
  \end{bmatrix}$$

\noindent implies that

\begin{equation*}
    \rho\Big(C_t\int_0^\infty w_a e^{-r a} da\Big) = \frac{\splitfrac{\sqrt{
    \begin{aligned}
    \Big(\int_0^\infty (c_{11}w^1_s-c_{22}w^2_s) e^{-r a} da\Big)^2\\+4\Big(c_{12}\int_0^\infty w^1_s e^{-r a} da\Big)\Big(c_{21}\int_0^\infty w^2_s e^{-r a} da\Big)
    \end{aligned}
    }}{+ \int_0^\infty (c_{11}w^1_s+c_{22}w^2_s) e^{-r a} da}}{2}
\end{equation*}

Therefore, the relationship between the overall reproduction number $\gamma_t$ and epidemic growth rate $r$ is given by

$$
\frac{1}{R_t}=\frac{\splitfrac{\sqrt{
    \begin{aligned}
    \Big(\int_0^\infty (c_{11}w^1_s-c_{22}w^2_s) e^{-r a} da\Big)^2\\+4\Big(c_{12}\int_0^\infty w^1_s e^{-r a} da\Big)\Big(c_{21}\int_0^\infty w^2_s e^{-r a} da\Big)
    \end{aligned}
    }}{+ \int_0^\infty (c_{11}w^1_s+c_{22}w^2_s) e^{-r a} da}}{c_{11}+c_{22} + \sqrt{(c_{11}-c_{22})^2+4c_{12}c_{21}}}
$$

\noindent More particularly if 

$$
C_t=c\begin{bmatrix}
    2 & 1 \\
    1 & 1
  \end{bmatrix}
$$

\noindent then this relationship becomes

\begin{align*}
    \frac{1}{R_t} &=\frac{\int_0^\infty (2cw^1_s+cw^2_s) e^{-r a} da + \sqrt{
    4c^2\Big(\int_0^\infty w^1_s e^{-r a} da\Big)^2+c^2\Big(\int_0^\infty w^2_s e^{-r a} da\Big)^2
    }}{3c + c\sqrt{5}}\\
    &=\frac{\int_0^\infty (2w^1_s+w^2_s) e^{-r a} da + \sqrt{
    4\Big(\int_0^\infty w^1_s e^{-r a} da\Big)^2+\Big(\int_0^\infty w^2_s e^{-r a} da\Big)^2
    }}{3 + \sqrt{5}}
\end{align*}

\end{document}